\documentclass[twoside,twocolumn,9pt]{article}
\usepackage{extsizes}
\usepackage[super,sort&compress,comma]{natbib} 
\usepackage[version=3]{mhchem}
\usepackage[left=1.5cm, right=1.5cm, top=1.785cm, bottom=2.0cm]{geometry}
\usepackage{balance}
\usepackage{times,mathptmx}
\usepackage{sectsty}
\usepackage{graphicx} 
\usepackage{lastpage}
\usepackage[format=plain,justification=raggedright,singlelinecheck=false,font={stretch=1.125,small,sf},labelfont=bf,labelsep=space]{caption}
\usepackage{float}
\usepackage{fancyhdr}
\usepackage{fnpos}
\usepackage[english]{babel}
\usepackage{array}
\usepackage{droidsans}
\usepackage{charter}
\usepackage[T1]{fontenc}
\usepackage[usenames,dvipsnames]{xcolor}
\usepackage{setspace}
\usepackage[compact]{titlesec}
\usepackage{amssymb}

\usepackage{epstopdf}

\definecolor{cream}{RGB}{222,217,201}

\begin{document}

\pagestyle{fancy}
\thispagestyle{plain}
\fancypagestyle{plain}{

	\renewcommand{\headrulewidth}{0pt}
}

\makeFNbottom
\makeatletter
\renewcommand\LARGE{\@setfontsize\LARGE{15pt}{17}}
\renewcommand\Large{\@setfontsize\Large{12pt}{14}}
\renewcommand\large{\@setfontsize\large{10pt}{12}}
\renewcommand\footnotesize{\@setfontsize\footnotesize{7pt}{10}}
\makeatother

\renewcommand{\thefootnote}{\fnsymbol{footnote}}
\renewcommand\footnoterule{\vspace*{1pt}%
\color{cream}\hrule width 3.5in height 0.4pt \color{black}\vspace*{5pt}} 
\setcounter{secnumdepth}{5}

\makeatletter 
\renewcommand\@biblabel[1]{#1}            
\renewcommand\@makefntext[1]%
{\noindent\makebox[0pt][r]{\@thefnmark\,}#1}
\makeatother 
\renewcommand{\figurename}{\small{Fig.}~}
\sectionfont{\sffamily\Large}
\subsectionfont{\normalsize}
\subsubsectionfont{\bf}
\setstretch{1.125} 
\setlength{\skip\footins}{0.8cm}
\setlength{\footnotesep}{0.25cm}
\setlength{\jot}{10pt}
\titlespacing*{\section}{0pt}{4pt}{4pt}
\titlespacing*{\subsection}{0pt}{15pt}{1pt}

\fancyfoot{}
\fancyfoot[RO]{\footnotesize{\sffamily{1--\pageref{LastPage} ~\textbar  \hspace{2pt}\thepage}}}
\fancyfoot[LE]{\footnotesize{\sffamily{\thepage~\textbar\hspace{3.45cm} 1--\pageref{LastPage}}}}
\fancyhead{}
\renewcommand{\headrulewidth}{0pt} 
\renewcommand{\footrulewidth}{0pt}
\setlength{\arrayrulewidth}{1pt}
\setlength{\columnsep}{6.5mm}
\setlength\bibsep{1pt}

\makeatletter 
\newlength{\figrulesep} 
\setlength{\figrulesep}{0.5\textfloatsep} 

\newcommand{\topfigrule}{\vspace*{-1pt}%
\noindent{\color{cream}\rule[-\figrulesep]{\columnwidth}{1.5pt}} }

\newcommand{\botfigrule}{\vspace*{-2pt}%
\noindent{\color{cream}\rule[\figrulesep]{\columnwidth}{1.5pt}} }

\newcommand{\dblfigrule}{\vspace*{-1pt}%
\noindent{\color{cream}\rule[-\figrulesep]{\textwidth}{1.5pt}} }

\makeatother

\twocolumn[
	\begin{@twocolumnfalse}
		\vspace{3cm}
		\sffamily
		\begin{tabular}{m{4.5cm} p{13.5cm} }
			 & \noindent\LARGE{\textbf{Interlayer interaction and related properties of
	bilayer hexagonal boron nitride: \textit{ab initio} study}} \\
	\vspace{0.3cm} & \vspace{0.3cm} \\

		& \noindent\large{Alexander V. Lebedev,$^{\ast}$\textit{$^{a}$} Irina V. Lebedeva,\textit{$^{b}$} Andrey A. Knizhnik\textit{$^{a,c}$} and Andrey M. Popov\textit{$^{d}$}} \\
	 & \noindent\normalsize{
	Principal characteristics of interlayer interaction and relative motion of hexagonal boron nitride (h-BN) layers are investigated by the first-principles method taking into account van der Waals interactions. Dependences of the interlayer interaction energy on relative translational displacement of h-BN layers (potential energy surfaces) are calculated for two relative orientations of the layers, namely, for the layers aligned in the same direction and in  the opposite directions upon relative rotation of the layers by 180 degrees. It is shown that the potential energy surfaces of  bilayer h-BN can be approximated by the first Fourier components determined by symmetry. As a result, a wide set of physical qualintities describing relative motion of h-BN layers aligned in the same direction including barriers to their relative sliding and rotation, shear mode frequency and shear modulus are determined by a single parameter corresponding to the roughness of the potential energy surface, similar to bilayer graphene. The properties of h-BN layers aligned in the opposite directions are described by two such parameters. The possibility of partial and full dislocations in stacking of the layers is predicted for h-BN layers aligned in the same and opposite directions, respectively.  The extended two-chain Frenkel-Kontorova model is used to estimate the width and formation energy of these dislocations on the basis of the calculated potential energy surfaces.
}

 \end{tabular}
 \end{@twocolumnfalse} \vspace{0.6cm}
]

\renewcommand*\rmdefault{bch}\normalfont\upshape
\rmfamily
\section*{}
\vspace{-1cm}


\footnotetext{\textit{$^{a}$~Kintech Lab Ltd., Moscow 123182, Russia, 
E-mail: allexandrleb@gmail.com}}
\footnotetext{\textit{$^{b}$~Nano-Bio Spectroscopy Group and ETSF Scientific Development Centre, Departamento
		de F\'{i}sica de Materiales, Universidad del Pa\'{i}s Vasco UPV/EHU, San Sebastian E-20018, Spain, E-mail:
liv\_ira@hotmail.com}}
\footnotetext{\textit{$^{c}$~National Research Centre "Kurchatov Institute", Moscow 123182, Russia}}
\footnotetext{\textit{$^{d}$~Institute for Spectroscopy of Russian Academy of Sciences, Troitsk, Moscow 142190, Russia, E-mail: popov-isan@mail.ru }}


\section{Introduction}
Though hexagonal boron nitride (h-BN)  is one of the main components of nanodevices along with graphene, less
is known about its physical properties. Nevertheless, a large number of  phenomena 
\cite{Rong1993,
Gan2003,
Warner2009,
Dienwiebel2004,
Dienwiebel2005,
Filippov2008,
Alden2013,
Zheng2008,
Lebedeva2010,
Lebedeva2011a,
Bistritzer2010,
Bistritzer2011,
Poklonski2013,
Lam2009,
Qian2012,
Zheng2012,
Paulla2013,
Brown2012,
Butz2013,
Lin2013,
Yankowitz2014,
Hattendorf2013,
San-Jose2014,
Lalmi2014,
Benameur2015,
Gong2013,
Koshino2013}
related to interaction of 2D layers discovered first for bilayer and few-layer graphene can also be expected for hexagonal boron nitride and can be useful for development of new applications. These phenomena include among others observation of Moir\'{e} patterns upon relative rotation of graphene layers,\cite{Rong1993,Gan2003,Warner2009} atomic-scale slip-stick motion of a graphene flake attached to STM tip on a graphene surface,\cite{Dienwiebel2004,Dienwiebel2005,Filippov2008} self-retracting motion of the
layers at their telescopic extension,\cite{Zheng2008} diffusion and drift of a graphene flake on a graphene surface via rotation to incommensurate states \cite{Lebedeva2010,Lebedeva2011a} and observation of dislocations in stacking of graphene layers.\cite{Alden2013, Brown2012,
Butz2013, Lin2013, Yankowitz2014, Hattendorf2013, San-Jose2014, Lalmi2014, Benameur2015,
Gong2013} The latter takes place when different parts of the system experience different relative shifts between the layers at the atomic scale and the system gets divided into a number of commensurate domains separated by incommensurate boundaries. Numerous experimental and theoretical studies have revealed that such dislocations can be used to tune electronic\cite{Hattendorf2013, San-Jose2014, Lalmi2014, Benameur2015, Koshino2013} and optical\cite{Gong2013} properties of graphene and thus hold great promise for application in nanoelectronic devices. Variation of electronic properties of graphene layers upon their relative in-plane displacement\cite{Bistritzer2010,Bistritzer2011,Poklonski2013} has already been proposed as a basis for various nanosensors.\cite{Poklonski2013,Lam2009,Qian2012,Zheng2012,Paulla2013} Understanding of interlayer interaction in h-BN can result in equally interesting observations and developments. It is not suprising that considerable efforts are being made to investigate structural and energetic characteristics of
h-BN.
\cite{Pease1950,
	Pease1952,
	Lynch1966,
	Solozhenko1995,
	Solozhenko1997,
	Solozhenko2001,
	Paszkowicz2002,
	Bosak2006,
	Fuchizaki2008,
	Nemanich1981,
	Reich2005,
	Marini2006,
	Li2009,
	Zhi2009,
	Warner2010,
	Nag2010,
	Albe1997,
	Kern1999,
	Ohba2001,
	Janotti2001,
	Rydberg2003,
	Liu2003,
	Tohei2006,
	Ooi2006,
	Serrano2007,
	Hamdi2010,
	Marom2010,
	Constantinescu2013,
Hsing2014}

The structure and elastic properties of the ground state of bulk h-BN have been studied experimentally in
detail.  The interlayer distance,\cite{Pease1950, Pease1952, Lynch1966, Solozhenko1995, Solozhenko1997,
Solozhenko2001, Paszkowicz2002, Bosak2006, Fuchizaki2008} bulk
modulus,\cite{Solozhenko1995,Solozhenko1997,Bosak2006,Fuchizaki2008} shear modulus,\cite{Bosak2006} shear mode
frequency\cite{Nemanich1981,Reich2005} ($E_{2g}$ mode with adjacent layers sliding rigidly in the opposite
in-plane directions) and frequency of relative out-of-plane vibrations ($B_{1g}$ ZO mode with adjacent layers
sliding rigidly towards and away from each other) \cite{Marini2006} have been measured and stacking of the
layers has been established. \cite{Pease1950,Li2009} The interlayer distance has been also determined for
few-layer h-BN,\cite{Zhi2009,Warner2010,Nag2010} while for bilayer h-BN, only stacking is known from the
experimental studies.\cite{Warner2010} Rotational stacking faults have been observed for few-layer
h-BN.\cite{Warner2010} However, no experimental data on the interlayer binding energy, corrugation of the
potential surface of interlayer interaction energy and barriers to in-plane relative displacement of the
layers are available neither for bilayer nor for few-layer or bulk h-BN. 

State-of-the-art first-principles methods have been applied to calculate the interlayer distance,\cite{Albe1997,
	Kern1999,
	Ohba2001,
	Janotti2001,
	Rydberg2003,
	Liu2003,
	Tohei2006,
	Marini2006,
	Ooi2006,
	Serrano2007,
	Hamdi2010,
	Nag2010,
	Marom2010,
Constantinescu2013}
interlayer binding energy,\cite{Nag2010,Marom2010,Rydberg2003}  bulk
modulus,\cite{Albe1997,Kern1999,Janotti2001,Ohba2001,Rydberg2003,Tohei2006,Ooi2006,Hamdi2010} shear
modulus,\cite{Hamdi2010} shear mode frequency \cite{Ohba2001,Serrano2007,Hamdi2010} and frequency of
out-of-plane relative vibrations  \cite{Kern1999,Ohba2001,Marini2006,Serrano2007,Hamdi2010} for bulk h-BN.
Much less theoretical data are available for bilayer h-BN, including the interlayer
distance,\cite{Rydberg2003,Nag2010,Marom2010,Constantinescu2013,Hsing2014} interlayer binding energy,
\cite{Rydberg2003,Nag2010,Marom2010,Hsing2014} bulk modulus and frequency of out-of-plane
vibrations.\cite{Nag2010} The dependence of the interlayer binding energy on the layer stacking has been
analyzed both for bulk \cite{Liu2003,Ooi2006,Constantinescu2013} and bilayer h-BN
\cite{Marom2010,Constantinescu2013,Gao2015} and the barriers to relative sliding of the layers have been
calculated.\cite{Marom2010,Constantinescu2013,Gao2015} However, the results of these calculations on relative
in-plane motion of h-BN layers are rather contradictory and obtained with the aim to study the balance between
van der Waals and electrostatic forces without relation to measurable physical quantities.

Considerably more detailed theoretical and experimental data are available for interaction and relative motion
of graphene layers. The approximation \cite{Verhoeven2004} of the interaction energy of a single carbon atom
in a graphene flake with a graphite layer by the first Fourier components determined by graphene symmetry has
been used to derive a simple expression for the potential surface of interaction energy of graphene
layers.\cite{Lebedeva2010,Lebedeva2011a,Reguzzoni2012} The latter has been confirmed both by density
functional theory calculations with the dispersion correction (DFT-D) and calculations using empirical
potentials with very good accuracy.\cite{Lebedeva2011a,Lebedeva2011,Popov2012} According to this expression,
the shape of the potential energy surface at a given interlayer distance is determined by a single parameter
and thus links different physical quantities related to relative motion of the layers such as barriers to
relative sliding and rotation, shear mode frequencies and shear modulus. Moreover, the approximation allowed
to apply the Frenkel-Kontorova model extended to the case of two interacting structures with a slightly
different lattice constant \cite{Bichoutskaia2006} to calculate the width and formation energy of dislocations in stacking of graphene layers and to consider the commensurate-incommensurate phase
transition in bilayer graphene upon stretching of one of the layers. \cite{Popov2011} Using the approximation,
the barrier to relative motion of the layers in bilayer graphene was estimated on the basis of experimentally
measured values of the shear mode frequency and width of dislocations in the
layer stacking  to be 1.7 meV/atom (Ref.~\cite{Popov2012}) and 2.4 meV/atom (Ref.~\cite{Alden2013}),
respectively (note that all energies for bilayers in the present paper are represented in meV per atom in the
top (adsorbed) layer so that they are more or less equal to the exfoliation energy of the bulk material per
atom and adsorption energy of a flake on the 2D layer per atom of the flake). These values are in good
agreement with those predicted by theory: $\sim$1 meV/atom (Ref.~\cite{Kolmogorov2005}, DFT calculations for
graphite within local density approximation (LDA) and generalized gradient approximation (GGA)), 2.6 meV/atom
(Ref.~\cite{Reguzzoni2012}), 1.82 meV/atom (Ref.~\cite{Aoki2007}, LDA calculations for bilayer graphene), $1 -
1.5 $ meV/atom (Ref.~\cite{Ershova2010}, DFT-D calculations for polycyclic aromatic hydrocarbons adsorbed on
graphene),  2.6 meV/atom (Ref.~\cite{Reguzzoni2012}), 2.07 meV/atom (Ref.~\cite{Lebedeva2011}, DFT-D
calculations for bilayer graphene), 1 meV/atom (Ref.~\cite{Reguzzoni2012}),  1.92 meV/atom
(Ref.~\cite{Lebedeva2011}, calculations for bilayer graphene using the vdW-DF functional \cite{Dion2004}). It
should be also noted that though there is a considerable discrepancy in the experimental estimates of the
barrier to relative motion of graphene layers, its relative value is even less than that for the exfolitation
energy of graphite ranging from $31\pm2$ to $52\pm5$ meV/atom
(Refs.~\cite{Girifalco1956,Benedict1998,Zacharia2004,Liu2012}).

The aim of the present paper is to predict experimentally measurable physical quantities associated with the
potential surface of interaction energy of h-BN layers. We use the vdW-DF2 approach  \cite{Lee2010} and
adequacy of our calculations is confirmed by the agreement in the interlayer interaction energy for symmetric
stackings of  h-BN layers with the values obtained recently by local second-order M{\o}ller-Plesset
perturbation theory (LMP2),\cite{Constantinescu2013} which is a high-level \textit{ab initio} method that,
different from DFT,  fully describes vdW interactions. Analogously to the expression for the potential surface
of interaction energy between graphene layers, simple expressions describing potential energy surfaces of h-BN
layers aligned in the same and opposite directions are introduced with the parameters fitted to the results of the
calculations. Based on these expressions a set of physical quantities related to the interlayer interaction
and relative motion of h-BN layers is estimated, including the barriers to relative motion and rotation of the
layers, shear mode frequencies and shear modulus for bilayer and bulk h-BN, length and formation energy of a
dislocation in stacking of the layers in bilayer h-BN. Implications of the
approximation introduced for the h-BN potential energy surface for further calculations of h-BN mechanical
properties and simulations of dynamical phenomena related to oscillation of h-BN layers are discussed.  The
results obtained are also important for understanding the interactions in recently discovered 
\cite{Dean2010,Ponomarenko2011,Britnell2012,Woods2014} graphene-h-BN
heterostructures (see also Ref. \cite{Geim2013} for a
review) showing the commensurate-incommensurate phase transition.\cite{Woods2014} The calculations
performed here represent a necessary intermediate stage (along with the previous calculations for bilayer
graphene \cite{Lebedeva2011,Popov2012}) before theoretical consideration of properties of such
heterostructures related to the interlayer interaction.

The paper is organized in the following way. Section 2 is devoted to first-principles calculations of elastic properties of a single h-BN layer and the potential
surface of interlayer interaction energy in bilayer and bulk h-BN. The estimates of physical
quantities associated with the interaction and relative motion of h-BN layers are obtained in section 3. Our conclusions are summarized in section 4.

\section{Methods and Results}
\subsection{DFT calculations of potential energy surfaces}

The DFT calculations have been performed using VASP code \cite{Kresse1996} with the non-local vdW-DF2
functional \cite{Lee2010} to take into account van der Waals (vdW) interactions. The basis set consists of
plane waves with the maximum kinetic energy of 600 eV. The interaction of valence electrons is described using
the projector augmented-wave method (PAW).\cite{Kresse1999} The periodic boundary conditions are applied to
the rectangular unit cell with 2 boron and 2 nitrogen atoms in each h-BN layer. The height of the simulation
cell is 20~\AA~for single-layer and bilayer h-BN and 6.66~\AA~for the bulk material. Integration over the
Brillouin zone is performed using the Monkhorst-Pack method \cite{Monkhorst1976} with the k-point grid of
$24\times 20 \times 1$ for single-layer and bilayer h-BN and $24\times 20 \times 4$ for bulk (in section 3,
axes $x$ and $y$ are chosen in the armchair and zigzag directions, respectively). Convergence with respect to
the number of k-points in the Brillouin zone and the maximum kinetic energy of plane waves was tested
previously for the energetic characteristics of bilayer graphene related to the interlayer
interaction.\cite{Lebedeva2011} The parameters chosen in the present study allow to evaluate these properties
within accuracy of 2 \%. The convergence threshold of the self-consistent field is $10^{-6}$ eV. 

The optimized bond length is $l = $1.455~\AA, in agreement with the experimental data for few-layer
\cite{Li2009,Zhi2009,Warner2010,Nag2010} and bulk h-BN \cite{Pease1950, Pease1952, Lynch1966, Solozhenko1995,
Solozhenko1997, Solozhenko2001, Paszkowicz2002, Bosak2006, Fuchizaki2008} and previous
calculations.\cite{Nag2010, Albe1997, Kern1999, Ohba2001, Janotti2001, Liu2003, Rydberg2003, Tohei2006,
Marini2006, Ooi2006, Serrano2007, Hamdi2010, Constantinescu2013} The effect of the interlayer interaction on
the structure of h-BN layers is neglected.\cite{Constantinescu2013}

The potential surface of interlayer interaction energy $U(x,y)$, i.e. the dependence of interlayer interaction
energy  $U$ on the relative displacement of two h-BN layers in the armchair ($x$) and zigzag directions ($y$)
at a fixed interlayer distance $d$, has been calculated by shifting the layers with respect to each other as
rigid. Account of deformations within the layers was shown to have a negligible effect on the potential
surface of interaction energy of carbon nanotube walls \cite{Belikov2004} and shells of carbon
nanoparticles.\cite{Lozovik2000, Lozovik2002} Taking into account the mirror planes and three-fold rotational
symmetry of a single h-BN layer, it is sufficient to explicitly calculate $U(x,y)$ only for 1/6 part of the
considered unit cell. The dependence $U(x,y)$ has been obtained for 78 points uniformly distributed on this
part of the unit cell.  The interlayer distance has been set at the experimental value of $d=3.33$~\AA~ for
bulk.\cite{Pease1950, Pease1952, Lynch1966, Solozhenko1995, Solozhenko1997, Solozhenko2001, Paszkowicz2002,
Bosak2006, Fuchizaki2008} The measurements for few-layer h-BN are less precise and include this value within
the error bars $3.25\pm0.10$~\AA~ (Ref.~\cite{Warner2010}). It should also be noted that the LMP2 calculations
\cite{Constantinescu2013} give the same interlayer distance of 3.34~\AA~ for bilayer and bulk h-BN.  Though
significant efforts have been made recently to improve description of the long-range vdW interactions in
DFT,\cite{Dion2004, Lee2010, Grimme2006, Grimme2010} available DFT methods still fail to predict the
equilibrium distance between 2D layers with sufficient accuracy.\cite{Lebedeva2011, Reguzzoni2012,
Constantinescu2013, Popov2013jcp} However, the magnitude of corrugations of the potential energy surface
\cite{Kolmogorov2005, Lebedeva2011, Reguzzoni2012} and barriers to relative motion of the layers
\cite{Lebedeva2011} depend exponentially on the interlayer distance. Therefore small deviations in the
equilibrium interlayer distance predicted by different DFT methods result in significant errors in the
properties related to interaction and relative motion of the layers.\cite{Lebedeva2011, Reguzzoni2012,
Constantinescu2013, Popov2013jcp} On the other hand, inclusion of vdW interactions almost does not affect the
potential energy surface at a given interlayer distance.\cite{Ershova2010, Lebedeva2011,
Reguzzoni2012,Marom2010,Constantinescu2013} Thus it is reasonable to study the potential energy surface at
the equilibrium interlayer distance known from the experiments rather than rely on the values provided by the DFT
methods.

\begin{figure*}
	\centering
	\includegraphics[width=\textwidth]{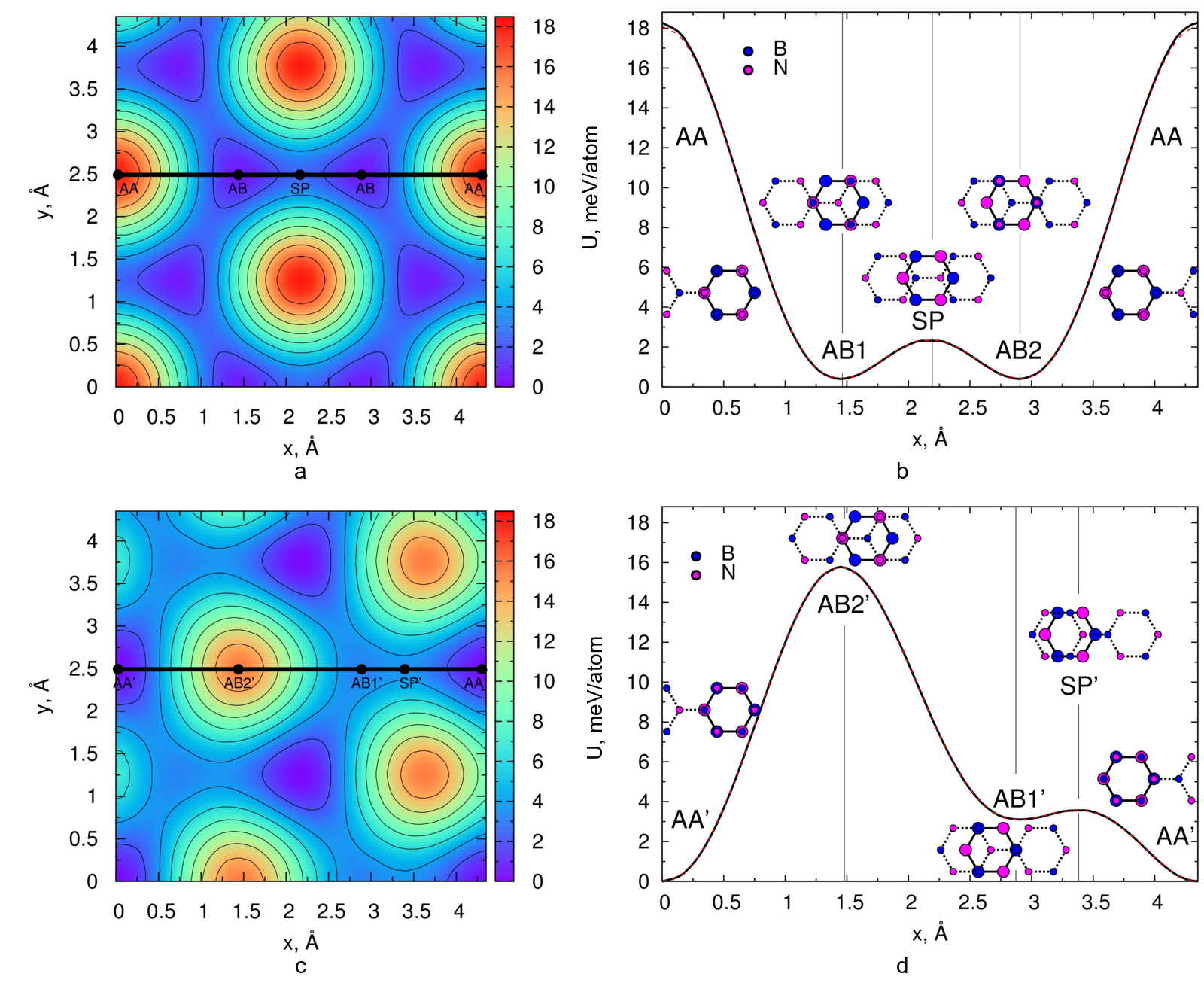}
	\caption{Calculated interlayer interaction energy of h-BN bilayer $U$ (in meV per atom of the top
	layer)  as a function of relative displacement of the layers in the armchair ($x$, in \AA) and zigzag
($y$, in \AA) directions at the interlayer distance of $d=$3.33~\AA: (a,b) h-BN layers aligned in the same
direction and (c,d) h-BN layers aligned in the opposite directions. The energy is given relative to the AA'
stacking. (b,d) Black solid lines correspond to the calculated dependences of interlayer interaction energy
$U$ on displacement $x$ in the armchair direction ($y = 0$) along the thick lines indicated in figures (a,c).
Curves approximated according to eqs.~\ref{eq_2} (b) and \ref{eq_3} (d) are shown by red dashed lines.
Structures of the symmetric stackings are indicated. Boron and nitrogen atoms are coloured in blue and
magenta, respectively.}
	\label{fig:pes}
\end{figure*}

\begin{table}[h]
	\caption{Calculated relative energies of symmetric stackings of h-BN bilayer with respect to the AA' stacking (in meV per atom of the top layer) for the interlayer distance of $d=$3.33~\AA. The LMP2 data are taken from Ref. \cite{Constantinescu2013}.  }
	\begin{tabular}{p{2cm}p{3cm}p{3cm}}
		\hline
		Stacking & vdW-DF2 & LMP2 \\\hline
		AB1' & 3.10 & 4.42 \\\hline
		AB2' & 15.77 & 16.50 \\\hline
		SP' & 3.57 &       \\\hline
		AB   & 0.40 & 0.24     \\\hline
		SP & 2.32 &      \\\hline
		AA   & 18.26 & 19.72 \\\hline\hline
	\end{tabular}\label{tab:energy_bilayer}
\end{table}

\begin{table}[h]
	\caption{Calculated relative energies of symmetric stackings of h-BN bulk with respect to the AA' stacking (in meV/atom) for the interlayer distance of $d=$3.33~\AA. The LMP2 data are taken from Ref. \cite{Constantinescu2013}. }

	\begin{tabular}{p{2cm}p{3cm}p{3cm}}
		\hline
		Stacking  & vdW-DF2 & LMP2 \\\hline
		AB1'  & 3.02      &  3.76     \\\hline
		AB2'  & 15.91    &  16.14    \\\hline
		SP'  &  3.57     &           \\\hline
		AB    & 0.38     &  0.44   \\\hline
		SP   &  2.39   &           \\\hline
		AA    & 18.47    &  19.89    \\\hline\hline
	\end{tabular}

	\label{tab:energy_bulk}
\end{table}

Let us now discuss the results of calculations of the potential energy surfaces for two orientations of h-BN
layers in the simulation cell (Fig.~\ref{fig:pes}). For h-BN layers aligned in the same direction
(Fig.~\ref{fig:pes}a and b), the potential surface of interlayer interaction energy is very similar to that of
graphene layers, in agreement with previous calculations.\cite{Constantinescu2013} The energy minima
correspond to the AB stacking in which nitrogen (boron) atoms of the top layer are located on top of boron
(nitrogen) atoms of the bottom layer and boron (nitrogen) atoms of the top layer are on top of hexagon
centers. It is clear that at this relative orienation of the layers all configurations corresponding to the
AB stacking with boron (AB2) or nitrogen (AB1) atoms of the top layer on top of the centers of the hexagons
are equivalent. Shifting the layers by half of the bond length in the armchair direction (towards the nearest minimum) brings the systems to the saddle-point (SP) stacking which corresponds to the barrier to
relative sliding of the layers. We find that this barrier is about 2 meV/atom (Tables~\ref{tab:energy_bilayer}
and \ref{tab:energy_bulk}), within the range reported for graphene \cite{Alden2013, Popov2012, Kolmogorov2005,
Reguzzoni2012, Aoki2007, Ershova2010, Lebedeva2011} and close to the LMP2 value $\sim$2.5 meV/atom for
h-BN.\cite{Constantinescu2013} Shifting the layers by one and half of the bond length more in the armchair
direction results in the AA stacking in which all atoms of the top layer are located on top of the equivalent
atoms of the bottom layer and the interlayer interaction energy reaches its maximum for the given interlayer
distance. The energy of the AA stacking relative to the AB one is about 18 meV/atom
(Tables~\ref{tab:energy_bilayer} and \ref{tab:energy_bulk}), close to the DFT data for graphene
\cite{Alden2013, Popov2012, Kolmogorov2005, Reguzzoni2012, Aoki2007, Ershova2010, Lebedeva2011} and for
bilayer\cite{Constantinescu2013} and bulk\cite{Liu2003,Ooi2006, Constantinescu2013} h-BN as well as LMP2
values of about 20 meV/atom \cite{Constantinescu2013} for bilayer and bulk h-BN.

The potential energy surface for h-BN layers aligned in the opposite directions (Fig.~\ref{fig:pes}c and d) is
rather different from those for graphene and h-BN layers aligned in the same direction, in agreement with
previous calculations.\cite{Marom2010,Constantinescu2013} In this case there are two types of inequivalent
AB' stackings. In the AB1(2)' stacking all boron (nitrogen) atoms of the top layer are on top of boron
(nitrogen) atoms of the bottom layer and the rest of atoms of the top layer are on top of hexagon centers.
Also the interlayer interaction energy reaches its maximum for the given orientation at the AB2' stacking,
while the local minima correspond to the AB1' stacking along with the AA' stacking in which all nitrogen and
boron atoms of the top layer are on top of boron and nitrogen atoms of the bottom layer, respectively. The
saddle-point stacking corresponding to the barrier to relative sliding of the layers lies on the straight path
between the nearest configurations corresponding to the AB1' and AA' but not in the middle (0.53~\AA~ from
AB1' and 0.92~\AA~ from AA') since the minima are of different depth.

All the DFT methods predict that the energies of the AB, AB1' and AA' stackings are rather close and differ
only by several meV/atom. However, the order of stability of this structures changes depending on the method
and chosen interlayer distance.\cite{Liu2003,Ooi2006,Nag2010, Marom2010,Constantinescu2013,Gao2015} Our
results indicate that the most stable
stacking both for h-BN bulk and bilayer is AA' (Tables~\ref{tab:energy_bilayer} and \ref{tab:energy_bulk}) 
and these results
are consistent with the experimental observations for bulk h-BN\cite{Pease1950} 
as well as with the LMP2 calculations \cite{Constantinescu2013} and DFT
calculations with account of nonlocal many-body dispersion (MBD) \cite{Gao2015} for bilayer and bulk h-BN.
The AB stacking is higher in energy only by 0.4 meV/atom and the relative energy of the AB1' stacking is 3
meV/atom, in agreement with the LMP2 results \cite{Constantinescu2013}
(Tables~\ref{tab:energy_bilayer} and \ref{tab:energy_bulk}) and the experimental data \cite{Warner2010} that
both the AA' and AB stackings are observed for bilayer h-BN. The AB1' minima in our case are very shallow and
are close in energy to the transition state SP' for relative sliding of the layers. The corresponding barrier
for transition from the AA' to AB1' stacking is 3.6 meV/atom, somewhat higher than in the case of h-BN layers
aligned in the same direction. 

\begin{table}[h]
	\caption{Calculated relative energies of symmetric stackings of h-BN bilayer with respect to the AA' stacking
	(in meV per atom of the top layer) for the interlayer distance of $d=$3.33~\AA~ with (vdW-DF2) and without (rPW86 (x) + LDA (c)) account of van der Waals interactions}
	\resizebox{0.5\textwidth}{!}{
		\begin{tabular}{lccc}
			\hline
			Stacking  & vdW-DF2 & rPW86 (x) + LDA (c)& difference \\\hline
			AB1' & 3.10   & 3.87 & -0.76      \\\hline
			AB2' & 15.77   & 15.30  & 0.48    \\\hline
			SP'  & 3.57   & 4.08  & 0.51   \\\hline
			AB   & 0.40   & 0.85  & -0.45     \\\hline
			SP   & 2.32   & 2.68  & 0.36   \\\hline
			AA   & 18.26   & 17.62  & 0.64    \\\hline\hline
		\end{tabular}
	}\label{tab:novdw}
\end{table}

It should be also mentioned that switching the vdW contribution off (using only LDA correlation and rPW86 exchange
\cite{Lee2010}) does not change the order of the symmetric stackings (Table~\ref{tab:novdw}). Their relative
energies with respect to the AA' stacking are changed by 0.4 - 0.8 meV/atom, which is within 5\% of the
magnitudes of corrugation of the potential energy surfaces both for h-BN layers aligned in the same and
opposite directions. The analogous conclusion was made previously for graphene bilayer
\cite{Ershova2010, Lebedeva2011,Reguzzoni2012} and h-BN bilayer \cite{Marom2010,Constantinescu2013,Gao2015} for a set of functionals with the dispersion correction.

Another observation is that the calculated relative energies of different stackings for bilayer and bulk h-BN
are nearly identical (Tables~\ref{tab:energy_bilayer} and \ref{tab:energy_bulk}), in agreement with previous
calculations for graphene.\cite{Lebedeva2011, Popov2012} The relative deviation is within 0.2 meV/atom, i.e.
about 1\% of the magnitudes of corrugation of the potential energy surfaces for h-BN layers aligned in the
same and opposite directions, and can be explained by interaction of non-adjacent layers in the bulk material.

\subsection{Approximation of potential energy surfaces}
To approximate the potential energy surface of bilayer h-BN we consider the top layer as adsorbed on the
bottom one and recall that the potential energy surface of an atom adsorbed on a 2D trigonal lattice can be
approximated by the first Fourier harmonics as \cite{Verhoeven2004}
\begin{equation} \label{eq_1}
	\begin{split}
		U = 0.5(U_0 + U_1(2\cos{(k_1 x)}\cos{(k_2 y)} +\cos{(2k_1 x)}+1.5)),
	\end{split}
\end{equation}
where $k_1 = 2\pi/(3l)$,  $k_2 = 2\pi/(\sqrt{3}l)$, $l$ is the bond length and the point $x=0$ and $y=0$
corresponds to the case when the atom is located on top of one of the lattice atoms. In the case of h-BN, we
should sum up interactions for boron atoms on the boron lattice (BB), nitrogen atoms on the nitrogen lattice
(NN), boron atoms on the nitrogen lattice (BN) and nitrogen atoms on the boron lattice (NB). In the latter two
contributions $U_{1NB}=U_{1BN}$ and $U_{0NB}=U_{0BN}$. Then for h-BN layers aligned in the same direction we
arrive at 
\begin{equation} \label{eq_2}
	\begin{split}
		U_\mathrm{I} = &U_{0,\mathrm{tot}} + 4.5 U_{1NB} + (U_{1NN}+U_{1BB}-U_{1NB})\times\\
			&\times(2\cos{(k_1 x)}\cos{(k_2 y)} +\cos{(2k_1 x)}+1.5),
	\end{split}
\end{equation}
where the point $x=0$ and $y=0$ corresponds to the AA stacking and $U_{0,\mathrm{tot}} = 2U_{0NB} + U_{0NN} +
U_{0BB} $.  It is seen that corrugations of the potential energy surface of h-BN layers aligned in the same direction are described by a single parameter $(U_{1NN}+U_{1BB}-U_{1NB})$, which determines all physical properties related to interlayer interaction for h-BN materials with adjacent layers aligned in the same direction.

For h-BN layers aligned in the opposite directions,
\begin{equation} \label{eq_3}
	\begin{split}
		U_\mathrm{II} = &U_{0,\mathrm{tot}} + 2.25(U_{1NN} + U_{1BB}) +(2U_{1NB}-0.5U_{1NN}-0.5U_{1BB}) \times \\ 
			 & \times (2\cos{(k_1 x)}\cos{(k_2 y)} +\cos{(2k_1 x)} + 1.5) -\\ 
	   &- \sqrt{3}(U_{1NN}-U_{1BB})\sin{(k_1 x)}(\cos{(k_2 y)}- \cos{(k_1 x)}),
	\end{split}
\end{equation}
where the point $x=0$ and $y=0$ corresponds to the AA' stacking. In this case corrugations of the potential energy surface are described by two parameters $(2U_{1NB}-0.5U_{1NN}-0.5U_{1BB})$ and $(U_{1NN}-U_{1BB})$ and these parameters are responsible for the properties of h-BN materials with adjacent layers aligned in the opposite directions.

The parameters of the approximation are found from the relative energies of the symmetric stackings AA', AB1', AB2' and AB as
\begin{equation} \label{eq_4}
	\begin{split}
		& U_{1NB} = \frac{E_\mathrm{AA'}- E_\mathrm{AB}}{4.5},\\
  & U_{1BB} = \frac{E_\mathrm{AB1'}- E_\mathrm{AA'} +9U_{1NB}}{4.5},\\
  & U_{1NN} = \frac{E_\mathrm{AB2'}- E_\mathrm{AA'} +9U_{1NB}}{4.5}.
	\end{split}
\end{equation}
The values obtained are $U_{1NB} = -0.0888$~meV/atom, $U_{1BB} = 3.328$~meV/atom and $U_{1NN} = 0.512$~meV/atom
and allow straightforward physical interpretation. The negative value of $U_{1NB}$ corresponds to the attraction
of ions of different sign, while positive values of $U_{1BB}$ and $U_{1NN}$ are related to repulsion of ions
of the same sign. The greater value of $U_{1BB}$ compared to $U_{1NN}$  is associated with the larger size of
the boron ion as compared to the nitrogen one.\cite{Huheey1993}

In addition to corrugations of the potential energy surface for h-BN layers aligned in the opposite directions
and relative energy of the co-aligned configuration, which are reproduced exactly, the approximation is very
accurate in relative energies of the SP and AA stackings for h-BN layers aligned in the same direction
(Fig.~\ref{fig:pes}b and d). The magnitude of corrugation for this orientation of the layers according to the
approximation is $E_\mathrm{AA} - E_\mathrm{AB} = 4.5 (U_{1NN}+U_{1BB}-U_{1NB}) = 17.68$~meV/atom, which is
only 0.18~meV/atom or 1\% smaller than the DFT value obtained. The barrier to relative sliding of the layers
aligned in the same directions according to the approximation is $E_\mathrm{SP} - E_\mathrm{AB} = 0.5
(U_{1NN}+U_{1BB}-U_{1NB}) = 1.96$~meV/atom. This is only 0.05~meV/atom or 2.6\% greater then the DFT value.
The standard deviation of the approximated potential energy surfaces from the ones obtained by the DFT
calculations for h-BN layers aligned in the same and opposite directions are 0.056~meV/atom and 0.014
meV/atom, respectively, which is within 0.3\% of their magnitudes of corrugation.

We should note that the possibility to accurately approximate potential energy surfaces by expressions
containing only the first Fourier harmonics has been previously demonstrated for interaction between graphene
layers\cite{Ershova2010,Lebedeva2011,Popov2012,Lebedeva2012} and between carbon nanotube walls, both for
infinite commensurate walls\cite{Belikov2004, Bichoutskaia2005, Bichoutskaia2009, Popov2009, Popov2012a} and
in the case where corrugations of the potential surface are determined by the contribution of edges
\cite{Popov2013} or defects.\cite{Belikov2004} Thus we can expect that analogous expressions can describe
potential energy surfaces for other layered materials with the van der Waals interaction between layers or for
translational motion of large molecules physically adsorbed on crystal surfaces. 

It should be also pointed out that the expression given by eq.~\ref{eq_2} for approximation of the potential energy surface for h-BN layers aligned in the same direction is exactly the same as the one for
graphene.\cite{Lebedeva2010,Lebedeva2011a,Popov2012,Reguzzoni2012} The calculated barrier to relative sliding of h-BN layers aligned in the same direction and the magnitude of corrugation of the potential energy surface (Tables~\ref{tab:energy_bilayer}
and \ref{tab:energy_bulk}) are within the ranges reported for graphene \cite{Alden2013, Popov2012, Kolmogorov2005,
Reguzzoni2012, Aoki2007, Ershova2010, Lebedeva2011}. Therefore it can be expected that these materials have very similar physical properties. This is confirmed in section 3 by comparison of physical quantities estimated for h-BN on the basis of eq.~\ref{eq_2} with the experimental data for graphene.

\subsection{Mechanical properties of single-layer h-BN} To get the Young modulus and Poisson ratio of a
single h-BN layer strains up to 0.8\% in the armchair direction and up to 0.5\% in the perpendicular zigzag
direction have been considered. For each size of the unit cell, positions of atoms within the cell have been
optimized using the quasi-Newton method \cite{Pulay1980} till the residual force of $6 \cdot 10^{-3}$~eV/\AA.
The energy dependence on the strain in the zigzag direction at a given  strain in the armchair direction has
been approximated by a parabola. The dependence of the energy corresponding to the parabola minimum on the
elongation in the armchair direction gives the elastic constant $k = 272.8 \pm 0.5$~J/m$^2$ or Young modulus
$Y = k/d = 819.2 \pm 1.5$~GPa, where $d = 3.33$~\AA~is the interlayer distance in bulk h-BN, while the
dependence of the position of the parabola minimum gives the Poisson ratio $\nu = 0.2011 \pm 0.0010$. These
values are in excellent agreement with the experimental data for bulk h-BN of $Y = 811$~GPa and $\nu = 0.21$
(Ref.~\cite{Bosak2006}) and are close to the DFT results for bulk h-BN of $Y = 860 - 900$~GPa, $\nu= 0.21 -
0.22$ (Ref.~\cite{Hamdi2010}),  $Y = 952$~GPa, $\nu= 0.18$ (Ref.~\cite{Ohba2001}) and for single-layer h-BN of
$k=279$~J/m$^2$, $\nu= 0.22$ (Ref.~\cite{Peng2012}) and 291~J/m$^2$, $\nu= 0.21$ (Ref.~\cite{Duerloo2012}).

\section{Properties related to interlayer interaction}

As shown in the previos section, potential surfaces of interlayer interaction energy for h-BN layers aligned
in the same and opposite directions can be approximated by relatively simple functions involving only one or two
independent parameters (eqs.~\ref{eq_2} and \ref{eq_3}), respectively. Therefore all properties of h-BN related to
interaction and relative motion of the layers are basically determined by these one or two parameters for the layers aligned in the same and opposite directions, respectively. Our DFT
calculations demonstrate that interaction of non-adjacent layers in bulk h-BN has a weak influence on the
potential energy surface so that the same parameters are responsible for physical behavior of bilayer,
few-layer and bulk h-BN.

\subsection{Shear mode frequency and shear modulus}

\begin{figure}
	\centering
	\includegraphics[width=\columnwidth]{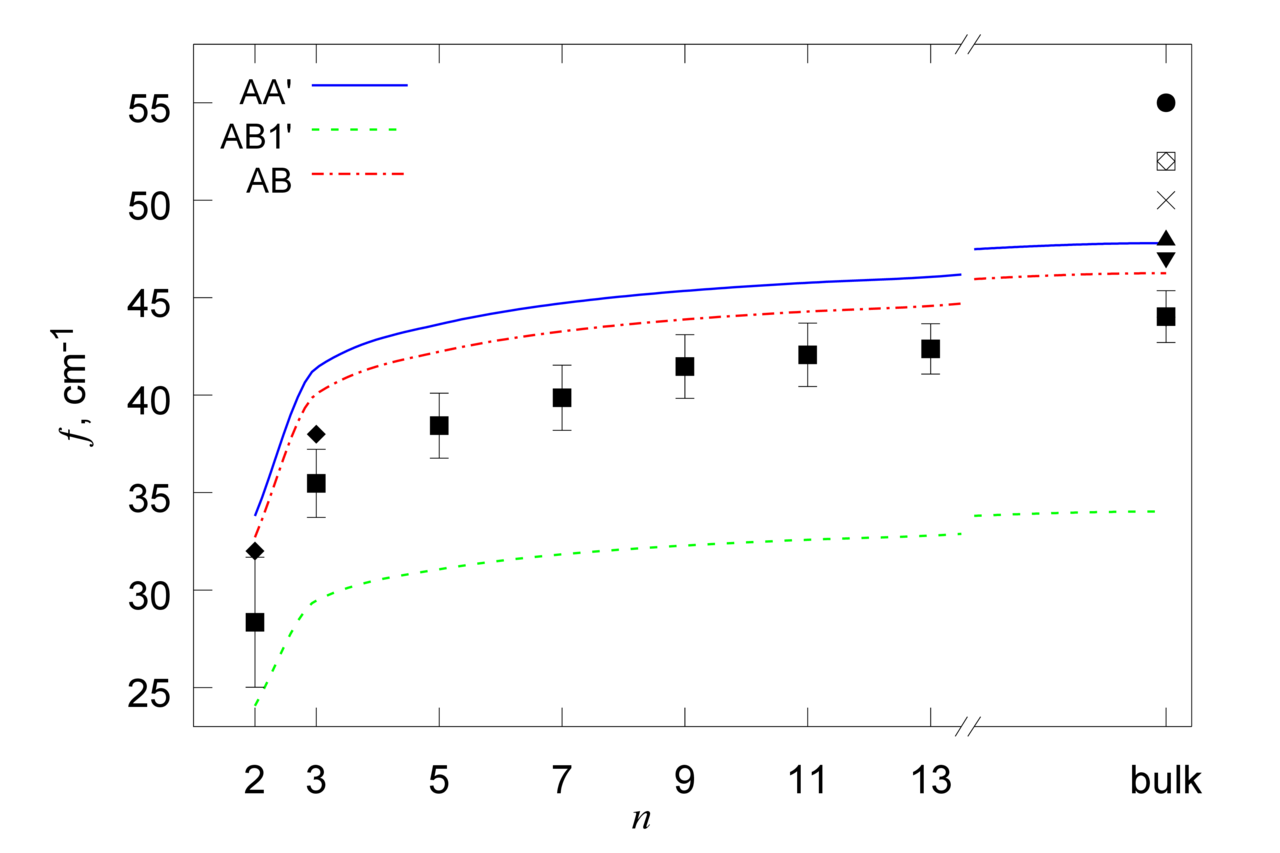}
	\caption{Calculated shear mode frequencies $f$ (in cm$^{-1}$) as functions of the number of layers $n$
    for h-BN crystals with different stacking of the layers: AA' --- blue solid line, AB1' --- green dashed
    line, AB --- red dash-dotted line. The experimental and theoretical data for bulk h-BN are included:
    $\square$ --- Ref.~\cite{Nemanich1981} (experiment),
    $\bullet$ --- Ref.~\cite{Reich2005} (experiment), $\times$ --- Ref.\cite{Ohba2001} (LDA), $\lozenge$ --- Ref.~\cite{Serrano2007} (LDA), $\blacktriangle$ --- Ref.\cite{Hamdi2010} (LDA), $\blacktriangledown$-- Ref.\cite{Hamdi2010} (revPBE). The experimental data for few-layer graphene and graphite are also shown for comparison: $\blacksquare$ --- Ref.~\cite{Boschetto2013} and $\blacklozenge$ --- Ref.~\cite{Tan2012}.}
	\label{fig:shear}
\end{figure}

\begin{figure}
	\centering
	\includegraphics[width=\columnwidth]{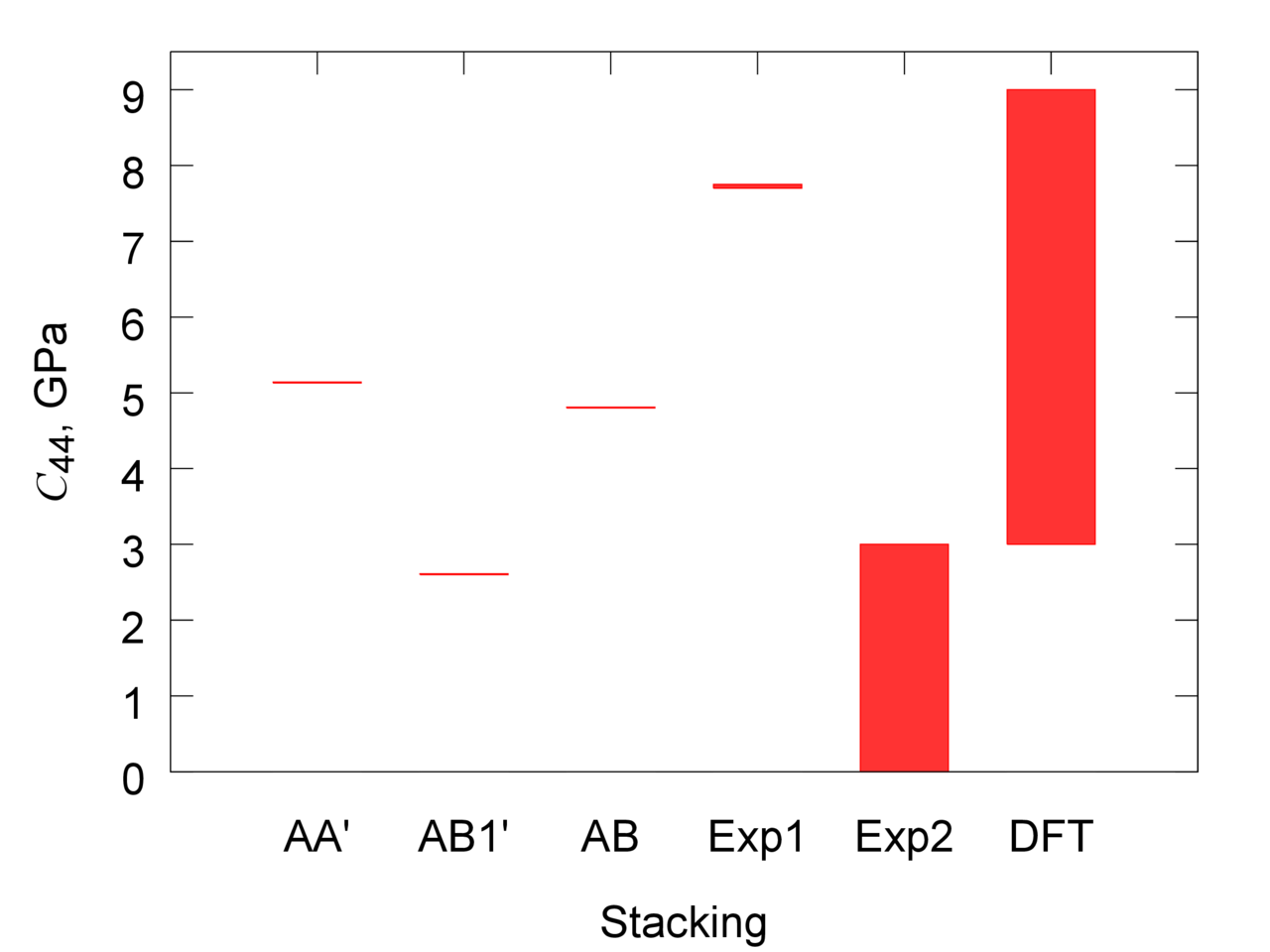}
	\caption{Calculated shear moduli $C_{44}$ (in GPa) for h-BN crystals (few-layer or bulk) with different stacking of the layers. The experimental data for bulk h-BN are taken from Ref. \cite{Bosak2006} (Exp1) and Ref. \cite{Duclaux1992} (Exp2). The range of values from the previous DFT calculations\cite{Hamdi2010} for bulk h-BN is indicated.}
	\label{fig:modulus}
\end{figure}

The frequency of the shear mode $E_{2g}$ in $n$-layer and
bulk ($n \to \infty$) h-BN, in which adjacent layers slide rigidly in the opposite in-plane directions, is
determined by the curvature of the potential energy surface in a given metastable state~\cite{Popov2012}
\begin{equation} \label{eq_6}
	\begin{split}
		f = \frac{1}{2\pi}\sqrt{\frac{n-1}{\mu}\frac{\partial^2 U}{\partial x^2}} = 
		\frac{1}{l}\sqrt{\frac{n-1}{\mu}U_{\mathrm{eff}}},
	\end{split}
\end{equation}
where $\mu = p(m_N+m_B)/4$ for $n = 2p$ and  $\mu = p(p+1)(m_N+m_B)/2(2p+1)$ for $n = 2p+1$, $m_B$ and $m_N$
are masses of boron and nitrogen atoms, respectively, $p$ is an integer, $l$ is the bond length and we denote
$U_{\mathrm{eff}} = (l/2\pi)^2 \partial^2 U/\partial x^2$. Curvatures of the potential energy surfaces in the
states corresponding to the AA', AB1' and AB stackings are described the parameters $U_{\mathrm{eff}} (AA') =
(U_{1NN}+U_{1BB}-4U_{1NB})/3 = 1.40$~meV/atom, $U_{\mathrm{eff}} (AB1') = (U_{1BB}+2U_{1NB}-2U_{1NN})/3 =
0.71$~meV/atom and $U_{\mathrm{eff}} (AB) = (U_{1NN}+U_{1BB}-U_{1NB})/3 = 1.31$~meV/atom. The corresponding
shear mode frequencies as functions of the number of layers $n$ are shown in Fig.~\ref{fig:shear}. The estimated shear mode
frequency for the ground-state AA' stacking of bulk h-BN is in good agreement with the
experimental data\cite{Nemanich1981,Reich2005} and results of previous DFT calculations\cite{Ohba2001,Serrano2007,Hamdi2010} (Fig.~\ref{fig:shear}). The calculated dependences of the shear mode frequencies for the AA' and AB stackings on the number of layers are similar to the experimental ones for graphene.~\cite{Tan2012,Boschetto2013}

The curvature of the potential energy surface also determines the shear modulus, which is expected to weakly
depend on the number of h-BN layers and can be estimated as
\begin{equation} \label{eq_7}
	\begin{split}
		C_{44} = \frac{d}{\sigma}\frac{\partial^2 U}{\partial x^2} = 
		\frac{4\pi^2 d U_{\mathrm{eff}}}{l^2 \sigma} ,
	\end{split}
\end{equation}
where $\sigma =3\sqrt{3}l^2/4$ is the area per atom in a h-BN layer and $d = 3.33$~\AA~is the interlayer
distance. The shear moduli obtained for different stackings of the layers are given in Fig.~\ref{fig:modulus}. It is seen that the result for the AA' stacking is within the range of the experimental\cite{Bosak2006,Duclaux1992} and theoretical values\cite{Hamdi2010} for bulk h-BN.

\subsection{Barrier to relative rotation of h-BN layers}
The same as for graphene,\cite{Lebedeva2010,Lebedeva2011a} the potential energy surface of h-BN layers rotated
to an incommensurate state should be smooth. Therefore the interaction energy in such states can be estimated
as an average of the interaction energy (eqs.~\ref{eq_2} and \ref{eq_3}) over in-plane
relative displacements of the layers in the commensurate states 
\begin{equation} \label{eq_5}
	\begin{split}
		U_{\mathrm{rot}} = &\langle U_\mathrm{I} \rangle_{x,y}  = \langle U_\mathrm{II} \rangle_{x,y} = \\
		     &U_{0,\mathrm{tot}}  + 1.5(U_{1NN}+U_{1BB}+2U_{1NB})
	\end{split}
\end{equation}

Thus the barriers to relative rotation of h-BN layers from metastable states corresponding to the AA', AB1'
and AB stackings are $E_{\mathrm{rot}} (AA') = U_{\mathrm{rot}} - U_{\mathrm{II}} (AA') =
1.5(U_{1NN}+U_{1BB}-4U_{1NB}) = 6.29$ meV/atom, $E_{\mathrm{rot}} (AB1') = -1.5U_{1NN}+U_{1BB}+3U_{1NB}= 2.29$
meV/atom and  $E_{\mathrm{rot}} (AB) = 1.5(U_{1NN}+U_{1BB})-U_{1NB}= 5.85$ meV/atom, respectively. The values
for the most stable AA' and AB stackings are on the order of the data for graphene $E_{\mathrm{rot}} \sim 5$
meV/atom (Ref.~\cite{Popov2012}) and 4 meV/atom (Ref.~\cite{Lebedeva2010,Lebedeva2011a}). The obtained
expressions for the potential energy surface (eqs.~\ref{eq_2} and \ref{eq_3}) can be also used to estimate the
barries to relative rotation of h-BN layers in configurations corresponding to Moir\'{e} patterns analogously
to studies performed for graphene.\cite{Yakobson}

\subsection{Dislocations}

\begin{figure*}
	\centering
	\includegraphics[width=\textwidth]{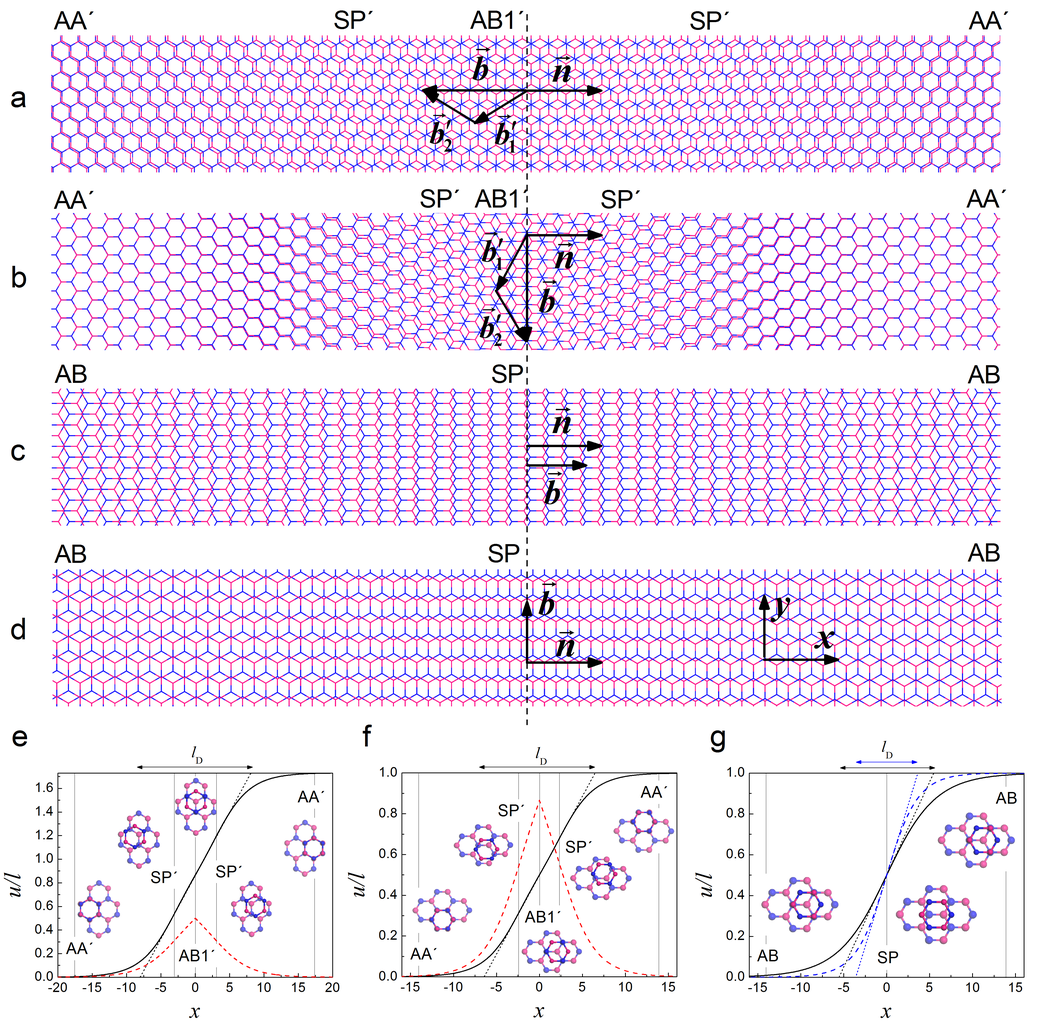}
	\caption{(a-d) Atomistic structures corresponding to full tensile (a) and shear (b) dislocations in h-BN bilayer with the layers aligned in the opposite directions and partial tensile (c) and shear dislocations (d) in h-BN bilayer with the layers aligned in the same direction. Boron and nitrogen atoms are coloured in blue and magenta, respectively. The directions of the unit vector $\vec{n}$ normal to the boundary between commensurate domains (black dashed line) and Burgers vector $\vec{b}$ are indicated. The magnitudes of the vectors are scaled up for clarity. (a,b) The directions of the Burgers vectors $\vec{b'}_1$ and $\vec{b'}_2$ corresponding to two straight pieces of the path of full dislocations ($\vec{b} = \vec{b'}_1+\vec{b'}_2$) are shown. (e-g) Displacements $u/l$ ($l=1.455$~\AA~is the bond length) of atoms as functions of their position $x$ (in nm) in the direction of the normal $\vec{n}$ to the boundary between commensurate domains for full tensile (e) and shear (f) dislocations in h-BN bilayer with the layers aligned in the opposite directions and for partial (g) tensile and shear dislocations in h-BN bilayer with the layers aligned in the same direction. (e,f) The displacements along the Burgers vector (black solid line) and in the perpendicular direction (red dashed line) are shown. (g) The displacement along the Burgers vector is shown. The displacement in the perpendicular direction is zero. The results for tensile (black solid line) and shear dislocations (blue dashed line) are given. The atomistic structures of symmetric stackings across the dislocations are included with their position shown by vertical gray lines. The characteristic widths $l_\mathrm{D}$ of the dislocations are indicated.}
	\label{fig:disl_struct}
\end{figure*}

Another manifestation of incommensurate states in two-dimensional layered materials is related to dislocations in stacking of the layers, which are observed when different parts of the system experience different relative shifts between the layers and correspond to incommensurate boundaries between commensurate domains. 
\cite{Alden2013, Brown2012, Butz2013, Lin2013, Yankowitz2014, Hattendorf2013, San-Jose2014, Lalmi2014, Benameur2015,
Gong2013} It has been demonstrated that such dislocations have a strong effect on the electronic\cite{Hattendorf2013, San-Jose2014, Lalmi2014, Benameur2015, Koshino2013} and optical\cite{Gong2013} properties of graphene. The analysis of structure of dislocations in few-layer graphene by transmission electron microscopy has already helped to get an experimental estimate of the barrier to relative displacement of the layers.\cite{Alden2013} Along with the properties discussed above the formation energy and width of dislocations in bilayer h-BN can be obtained from the potential surface of interlayer interaction energy.\cite{Popov2011}

To consider dislocations in bilayer h-BN we follow the formalism of the two-chain Frenkel-Kontorova model,\cite{Bichoutskaia2006} which was applied previously to study dislocations in double-walled carbon nanotubes\cite{Bichoutskaia2006, Popov2009} and graphene.\cite{Popov2011} Elementary dislocations in graphene correspond to partial dislocations with the Burgers vector $\vec{b}$ equal in magnitude to the bond length.\cite{Popov2011, Alden2013}  Similar dislocations ($b = l$, Fig.~\ref{fig:disl_struct}c,d) are also possible in h-BN layers aligned in the same direction as the potential surface of interlayer interaction in this case has degenerate energy minima AB1 and AB2 separated by the distance of one bond length $l$ (Fig.~\ref{fig:pes}a).  However, for h-BN layers aligned in the opposite directions, the AB1' energy minima are rather shallow and are much higher in energy than the AA' minima (Fig.~\ref{fig:pes}c). Therefore in this case the minimal possible Burgers vector is equal in magnitude to the lattice constant ($b = l \sqrt{3}$, Fig.~\ref{fig:disl_struct}a,b), i.e. the elementary dislocations are full. Both full dislocations in h-BN layers aligned in the opposite directions and partial dislocations in h-BN layers aligned in the same direction are considered below. It should be also noted that even for graphene theoretical estimates of the formation energy and width have been limited to the tensile dislocation with the Burger vector $\vec{b}$ normal to the boundary between commensurate domains,\cite{Popov2011} though experimentally dislocations with different angles between the Burgers vector and boundary between commensurate domains have been observed.\cite{Alden2013, Butz2013, Lin2013, Yankowitz2014, Hattendorf2013, Lalmi2014} Here we extend the Frenkel-Kontorova model to describe dislocations with the Burgers vector $\vec{b}$ at an arbitrary angle $\beta$ to the normal $\vec{n}$ to the boundary between commensurate domains. In particular, the results for shear dislocations (Fig.~\ref{fig:disl_struct}b,d), where the Burgers vector is parallel to the boundary between commensurate domains $\beta = \pm \pi/2$, are presented along with the results for tensile dislocations (Fig.~\ref{fig:disl_struct}a,c) with $\beta = 0$. 

In the limit of large systems and low density of dislocations it can be assumed that  dislocations are isolated. This means that the layers are commensurate at large distances  from the dislocation center. Correspondingly, the boundary conditions for the relative displacement  $\vec{u}$ of atoms of the layers can be formulated as  $\vec{u}=0$ at $x \to -\infty$ and $\vec{u}=-\vec{b}$ at $x \to +\infty$, where $x$ is the coordinate along the normal $\vec{n}$ to the boundary between commensurate domains and $\vec{b}$ is the Burgers vector. According to the Frenkel-Kontorova model,\cite{Bichoutskaia2006,Popov2011} the formation energy of isolated dislocations per unit length is determined by the sum of excessive elastic and van der Waals interaction energies originating from the deformation and incommensurability of the layers, respectively,
\begin{equation} \label{eq_7a}
	\begin{split}
		U_\mathrm{D} = \int\limits_{-\infty}^{+\infty} \left( \frac{1}{4} E |\vec{u}'_x|^2 + \frac{1}{4} G |\vec{u}'_y|^2 + V(\vec{u}) \right) \mathrm{d} x,
	\end{split}
\end{equation}
where the axis $y$ is chosen along the boundary between commensurate domains,  $\vec{u}' = \mathrm{d}\vec{u} /\mathrm{d}x$ describes the strain across the boundary, $V(\vec{u})$ is the van der Waals interaction energy per unit area relative to the deepest energy minimum for the given alignment of h-BN layers (the AB and AA' stackings for the layers aligned in the same and opposite directions, respectively), $E=k/(1-\nu^2)$ and $G=k/2(1+\nu)$ are the tensile and shear elastic constant of the layers, respectively, $k = Yd$ is the elastic constant under uniaxial stress and $\nu$ is the Poisson ratio.  Introducing the coefficient $(1-\nu^2)^{-1}$ in the tensile elastic constant $E$ we take into account that the tensile strain is kept zero in the direction along the boundary between commensurate domains (see also Ref.~\cite{Alden2013}). In the absence of an external load, one of the layers in the tensile dislocation is stretched in the direction accross the boundary, while the other one is compressed.\cite{Popov2011} Therefore in the perpendicular direction they tend to compress and stretch, respectively.  The interlayer interaction, however, keeps the layers commensurate in the latter direction, providing the vanishing tensile strain along the boundary in the regions far from boundary crossings. It should be also noted that the coefficient $(1-\nu^2)^{-1} = 1.04$ is only slightly different from unity and gives corrections to the formation energy and characteristic width of dislocations within 2\%.

The dislocation path, i.e. the dependence of the relative displacement $\vec{u}$ of h-BN layers on the coordinate $x$ in the direction perpendicular to the boundary between commensurate domains (Fig.~\ref{fig:disl_struct}e-g) that minimizes the formation energy $U_\mathrm{D}$ (eq.~\ref{eq_7a}), should satisfy the Euler-Lagrange equations 
\begin{equation} \label{eq_9a}
	\begin{split}
&\frac{\partial U_\mathrm{D}}{\partial u_x}=-\frac{1}{2} E u''_x+ \frac{\partial V(\vec{u})}{\partial u_x}=0, \\
&\frac{\partial U_\mathrm{D}}{\partial u_y}=-\frac{1}{2} G u''_y+ \frac{\partial V(\vec{u})}{\partial u_y}=0.
	\end{split}
\end{equation}
This means that the dislocation path is the same as the trajectory of a particle on the inverse potential energy surface $-V(\vec{u})$ with the fixed initial and final positions corresponding to the deepest energy minima of the original potential energy surface. It should be noted that the "particle mass" is anisotropic and equal to $G/2$ and $E/2$ in the directions along and across the boundary between commensurate domains, respectively. 

For partial dislocations in h-BN layers aligned in the same direction (Fig.~\ref{fig:disl_struct}c,d), the straight line between the adjacent AB minima (Fig.~\ref{fig:pes}a) corresponding to the minimum energy path satisfies eq.~\ref{eq_9a} and thus gives exactly the dislocation path. For full dislocations in h-BN layers aligned in the opposite directions (Fig.~\ref{fig:disl_struct}a,b), the solution of eq.~\ref{eq_9a} cannot be found analytically. Moreover, this solution should depend on the angle between the Burgers vector and the boundary between commensurate domains. However, it is clear that strong scattering introduced by rapidly changing potential at slopes of the AB2' hill (Fig.~\ref{fig:pes}c) should be avoided. Therefore we assume that the path of full dislocations in h-BN layers aligned in the opposite directions is also roughly described by the minimum energy path AA' -- AB1' -- AA' consisting of two straight lines at the angle $2 \pi/3$ to each other. More accurate approximations of the dislocation path give a correction to the formation energy within 10\%.

Integration of the Euler-Lagrange equations (eq.~\ref{eq_9a}) provides the analogue of the energy conservation law
\begin{equation} \label{eq_9}
	\begin{split}
		\frac{1}{4} k \phi^2 (\theta)|\vec{u}'|^2 = V(\vec{u}).
	\end{split}
\end{equation}
Here we use that the integration constant is zero for isolated dislocations\cite{Popov2011} and denote $\phi(\theta) =\sqrt{(E\cos^2 \theta + G \sin^2 \theta)/k}$, where $\theta$ is the angle between the direction $\vec{u}'$ of the dislocation path and normal $\vec{n}$ to the boundary between commensurate domains. Taking into account eq.~\ref{eq_9}, the formation energy of dislocations per unit length (eq.~\ref{eq_7a}) can then be expressed as
\begin{equation} \label{eq_7}
	\begin{split}
U_\mathrm{D} =  \int\limits_{-\infty}^{+\infty} \sqrt{k V(\vec{u})} \phi (\theta)|\vec{u}'| \mathrm{d} x.
	\end{split}
\end{equation}

\begin{figure}
	\centering
	\includegraphics[width=\columnwidth]{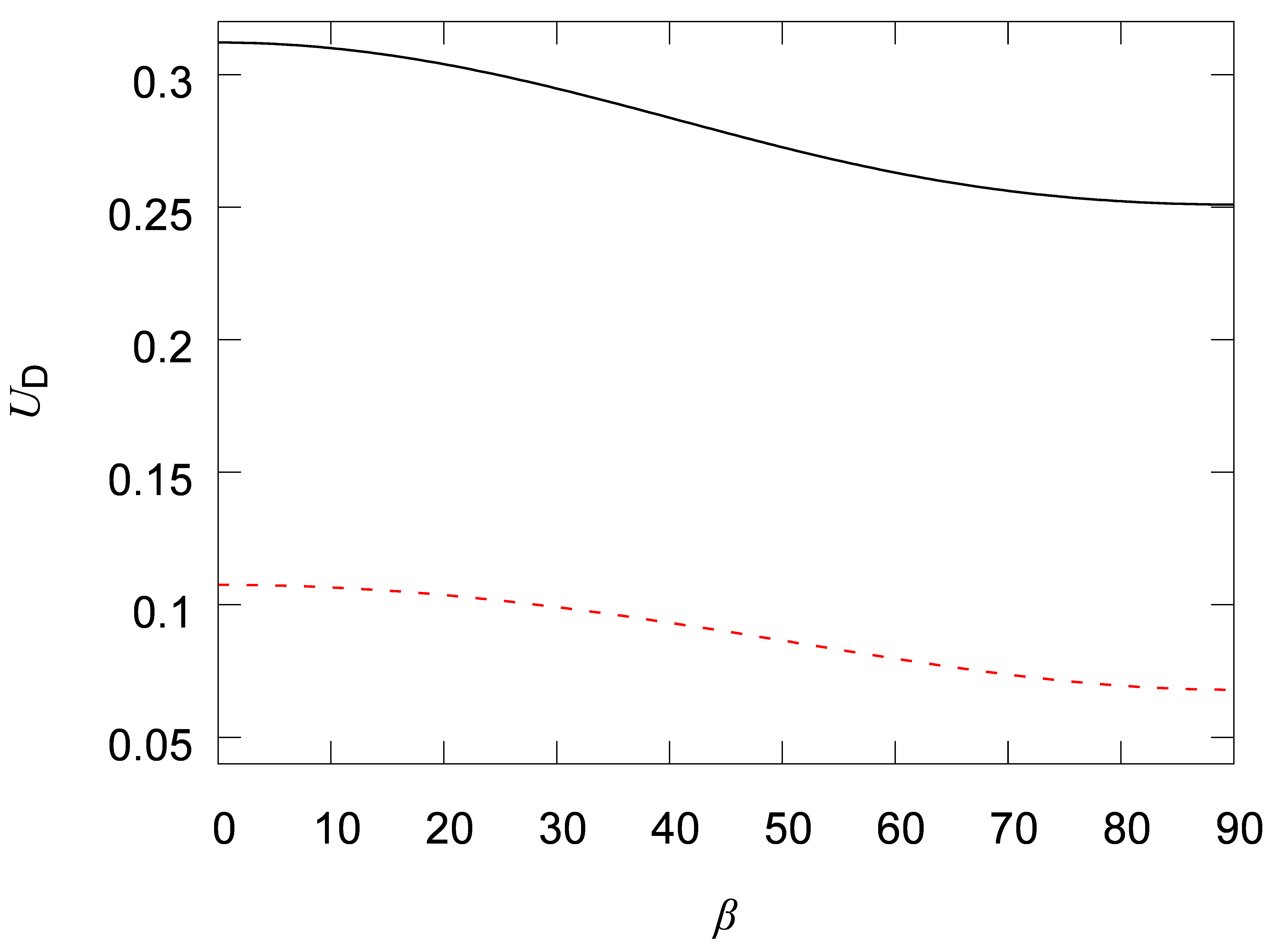}
	\caption{Calculated formation energy of dislocations $U_\mathrm{D}$ per unit width (in eV/\AA) as a function of angle $\beta$ (in degrees) between the Burgers vector $\vec{b}$ and normal $\vec{n}$ to the boundary between commensurate domains for a full dislocation (solid black line) in h-BN layers aligned in the opposite directions and a partial dislocation (dashed red line) in h-BN layers aligned in the same direction.}
	\label{fig:energy}
\end{figure}

\begin{figure}
	\centering
	\includegraphics[width=\columnwidth]{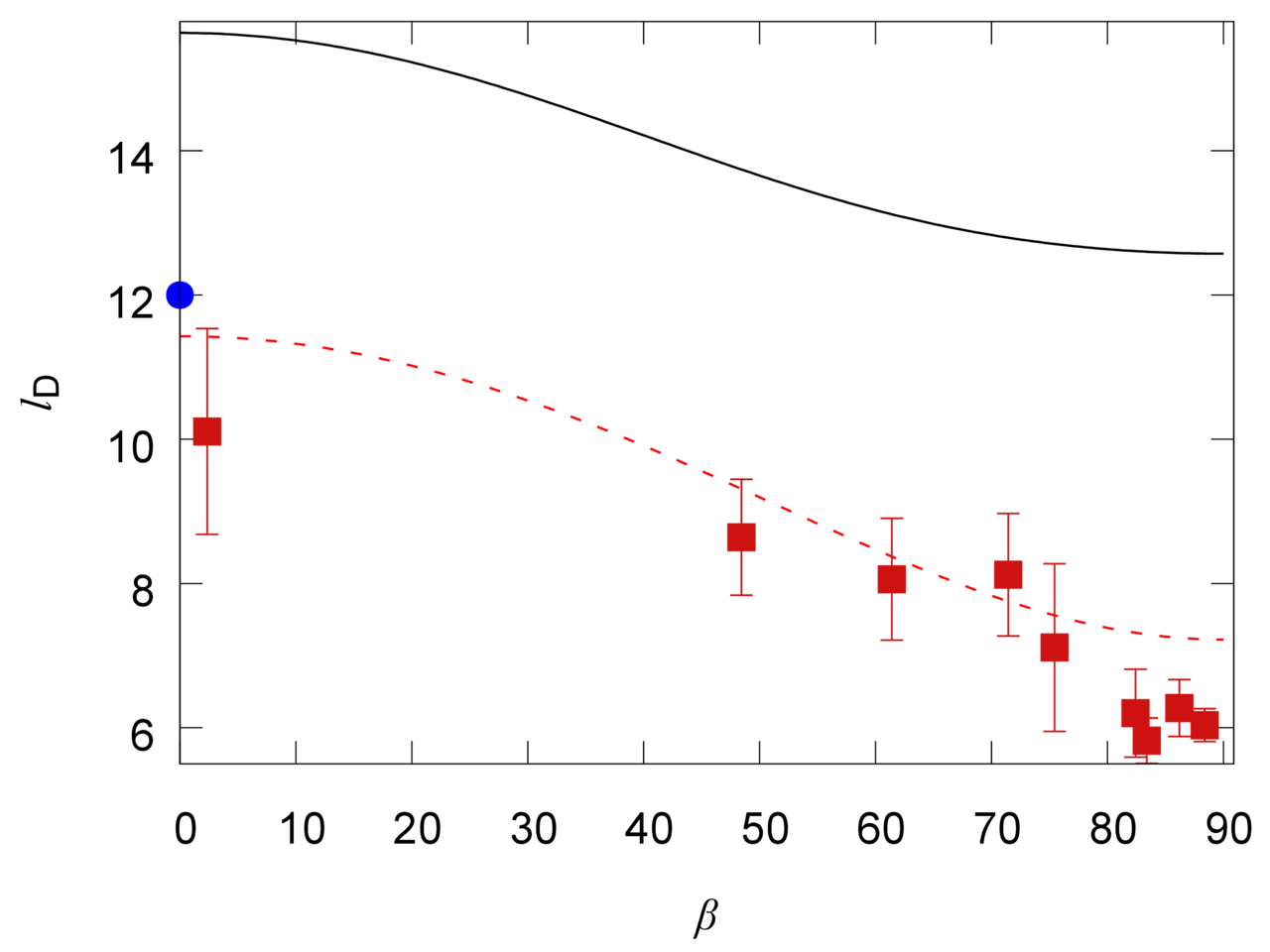}
	\caption{Calculated dislocation width $l_\mathrm{D}$ (in nm) as a function of angle $\beta$ (in degrees)
    between the Burgers vector $\vec{b}$ and normal $\vec{n}$ to the boundary between commensurate domains for
a full dislocation (solid black line) in h-BN layers aligned in the opposite directions and a partial
dislocation (dashed red line) in h-BN layers aligned in the same direction. The experimental data for graphene
from Ref.~\cite{Alden2013} are shown by squares ($\blacksquare$) with error bars. The DFT result\cite{Popov2011} for partial tensile dislocations in graphene is indicated by a circle ($\bullet$).}
	\label{fig:width}
\end{figure}

Let us first consider partial dislocations in h-BN layers aligned in the same direction (Fig.~\ref{fig:disl_struct}c,d). In these dislocations, atoms are displaced along straight lines, $\theta = \pi + \beta$ and $u$ changes from 0 to $b=l$. According to eq.~\ref{eq_7}, the formation energy of partial dislocations is 
\begin{equation} \label{eq_7b}
	\begin{split}
	U_\mathrm{D} (\beta) =\phi(\beta) \int\limits_0^l \sqrt{kV(u)} \mathrm{d} u,
	\end{split}
\end{equation}
where we approximate the interlayer interaction energy $V(u)$ along the minimum energy path on the basis of eq.~\ref{eq_2}. Fig.~\ref{fig:energy} demonstrates that the formation energy per unit length is maximal for tensile dislocations (Fig.~\ref{fig:disl_struct}c) and minimal for shear dislocations (Fig.~\ref{fig:disl_struct}d). Such dislocations correspond to the boundaries between commensurate domains in the zigzag and armchair directions, respectively. The estimated formation energies for h-BN layers aligned in the same direction (Fig.~\ref{fig:energy}) are on the order of the value predicted for partial tensile dislocations in bilayer graphene of 0.12~eV/\AA~(Ref. \cite{Popov2011}). 

Based on the approximation of the interlayer interaction energy along the minimum energy path by the cosine function and eq.~\ref{eq_9}, it was shown analytically that partial dislocations are described by a soliton with a relatively short incommensurate region separating commensurate domains.\cite{Popov2011} Though in the present paper we use eq.~\ref{eq_2} to approximate the potential energy surface of h-BN layers aligned in the same direction, the calculated path of partial dislocations is very close to the analytical solution (Fig.~\ref{fig:disl_struct}g). Since the slope of the dependence of displacement $u$ on position $x$ across the boundary is nearly constant around the dislocation center and close to the maximum value $|\vec{u}'|_\mathrm{max}$, the dislocation width can be defined in the way similar to the analytical solution
\begin{equation} \label{eq_10}
	\begin{split}
		l_\mathrm{D} (\beta) = \frac{l}{\left|\vec{u}' \right|_\mathrm{max}} =\phi(\beta) \frac{l}{2} \sqrt{\frac{k}{V_\mathrm{max}}},
	\end{split}
\end{equation}
where $V_\mathrm{max}$ is the barrier to relative sliding of the layers. It is clear that the dependence of the dislocation width on the angle $\beta$ between the Burgers vector and normal to the boundary between commensurate domains is the same as that of the formation energy (eq.~\ref{eq_7b}). In particular, the width of shear dislocations is smaller than the width of tensile ones (Fig.~\ref{fig:width}). The dependence of the formation energy and width of partial dislocations on the angle between the Burgers vector and boundary between commensurate domains is determined only by the Poisson ratio and should approximately hold for any hexagonal 2D crystals or heterostructures consisting of the layers with slightly different lattice constants where partial dislocations are possible. In particular, the calculated angular dependence of the dislocation width is in good agreement with experimentally measured values for graphene\cite{Alden2013,Lin2013,Yankowitz2014} ranging from 11 nm for tensile dislocations to 6 -- 7 nm for shear dislocations (Fig.~\ref{fig:width}). The calculated width of partial tensile dislocations is also on the order of the DFT estimate for graphene\cite{Popov2011} of 12 nm.

In the case of full dislocations in h-BN layers aligned in the opposite directions (Fig.~\ref{fig:disl_struct}a,b), it is needed to take into account two contributions from the straight pieces of the dislocation path at angles $\beta+\pi/6$ and $\beta-\pi/6$ to the normal $\vec{n}$ to the boundary between commensurate domains. This can be done by a simple substitution of $\phi(\beta)$ by $\phi(\beta- \pi/6)+\phi(\beta+ \pi/6)$ in eqs.~\ref{eq_7b} and \ref{eq_10}. All the conclusions made above for partial dislocations are also valid for full ones (Figs.~\ref{fig:energy} and \ref{fig:width}) with the only difference that full tensile and shear dislocations correspond to the boundaries between commensurate domains in the armchair and zigzag directions, respectively (Fig.~\ref{fig:disl_struct}a,b), opposite to the case of partial dislocations. It should be especially emphasized that according to Figs.~\ref{fig:disl_struct}e--g, both for partial and full dislocations the slope $|\vec{u}' (x)|$ is nearly constant for distances from the dislocation center comparable to the dislocation width. This gives the possibility of accurate measurements of the dislocation width and, consequently, of experimental estimation of the barrier to relative sliding of h-BN layers by transmission electron microscopy analysis of stacking at the boundary between commensurate domains in the same way as it was done for graphene.\cite{Alden2013}

\section{Conclusions}
DFT calculations of the potential surface of interlayer interaction energy at relative in-plane displacement
of h-BN layers have been performed using the vdW-DF2 functional.\cite{Lee2010} According to these results, the
ground state corresponds to the AA' stacking of the layers aligned in the opposite directions. However, the
optimal AB stacking for the layers aligned in the same direction is only 0.4 meV/atom less favourable, in
agreement with the experimental observation\cite{Warner2010} of both AB and AA' stackings for bilayer h-BN.
The magnitudes of corrugation of the potential energy surfaces are about 18 meV/atom and 16 meV/atom for the
layers aligned in the same and opposite directions, respectively. The barriers to relative motion of the
layers aligned in the same and opposite directions are 2 meV/atom and 3.6 meV/atom, respectively, in close
agreement with LMP2 calculations.\cite{Constantinescu2013} Almost identical results are obtained for bilayer
and bulk h-BN, indicating that interactions between non-adjacent layers have a weak effect on the
characteristics of the potential energy surface (1\% of the magnitude of corrugation). The contribution of vdW
interactions into these quantities is also small (within 5\% of the magnitude of corrugation) and comparable
to that in previous studies for h-BN \cite{Marom2010, Constantinescu2013}
and graphene.\cite{Ershova2010, Lebedeva2011,Reguzzoni2012}

It is shown that the calculated potential surfaces of interlayer interaction energy for h-BN layers aligned in the
same and opposite directions can be fitted by simple expressions containing only the first components of Fourier expansions determined by symmetry of the layers.
Analogous approximations have been suggested previously for graphene
bilayer\cite{Ershova2010,Lebedeva2011,Popov2012,Lebedeva2012} and double-walled carbon nanotubes
\cite{Belikov2004, Bichoutskaia2005, Bichoutskaia2009, Popov2009, Popov2012a, Popov2013} based both on DFT
calculations and semi-empirical potentials. Thus it can be expected that for other layered materials the
potential surface of interlayer interaction energy can be also reproduced closely by the first Fourier components determined by symmetry. 

Recently a concept of the registry index surface was proposed to predict qualitative features of the potential
surface of interlayer interaction energy of h-BN bilayer.\cite{Marom2010} The expressions introduced here give
the possibility to reproduce the shape and quantitative characteristics of the potential energy surface of
h-BN bilayer and can be useful for multiscale simulations of such phenomena as atomic-scale slip-stick motion
of a h-BN flake attached to STM tip on the h-BN surface, diffusion of a h-BN flake on the h-BN surface and
formation of stacking dislocations in h-BN bilayer (analogous to multiscale simulations of the phenomena
observed for graphene \cite{Dienwiebel2004, Dienwiebel2005, Filippov2008, Zheng2008, Lebedeva2010, Lebedeva2011a, Alden2013, Brown2012, Butz2013, Lin2013, Yankowitz2014, Hattendorf2013, San-Jose2014, Lalmi2014, Benameur2015, Gong2013}). 

According to the proposed approximation, a set of physical properties of h-BN materials related to relative
motion of the layers are determined by only one or two independent parameters for the layers aligned in the same and opposite directions, respectively. Namely, the shear mode
frequency, shear modulus and barrier to relative rotation of the layers have been estimated for bilayer and
bulk h-BN in different stackings. The possibility of partial and full dislocations in stacking of the layers is suggested for h-BN layers aligned in the same and opposite directions, respectively. Their width and formation energy are governed by the same parameters of the potential energy surface. 
In particular, it is shown that a simple link exists between the dislocation width $l_\mathrm{D}$ and barrier $V_\mathrm{max}$ to relative sliding of the layers 
$V_\mathrm{max} = C(\beta)/l^2_\mathrm{D}(\beta)$, where the coefficient $C(\beta)=\phi^2(\beta)kl^2/4$ is determined by the elastic properties of the layers.
Experimental measurements of the dislocation width, shear mode
frequency or shear modulus can help to refine the 
parameters of the functions approximating the potential
surfaces of interlayer interaction energy in h-BN and to obtain the barriers 
and other characteristics of these surfaces in the same way as
it was done for graphene.\cite{Alden2013,Popov2012}


\section{Acknowledgement}
Authors acknowledge the Russian Foundation of Basic Research (14-02-00739-a) and
computational time on the Supercomputing Center of Lomonosov Moscow State University and the Multipurpose
Computing Complex NRC "Kurchatov Institute". IL acknowledges financial support from the Marie Curie
International Incoming Fellowship (Grant Agreement PIIF-GA-2012-326435 RespSpatDisp), EU-H2020 project
"MOSTOPHOS" (n. 646259) and Grupos Consolidados del Gobierno Vasco (IT-578-13).

%
%
%
%
%

\bibliography{rsc} 
\bibliographystyle{rsc} 

\end{document}